\documentstyle[aps,pra]{revtex}

\newcommand{\bn}{\mbox{\boldmath$\nabla$}}
\newcommand{\bs}{\mbox{\boldmath$\sigma$}}

\begin{document}

\title{Relativistic corrections to the dipole polarizability of the ground
state of the molecular ion $\mbox{H}_2^+$}
\author{V.I.~Korobov}
\address{Institute of Theoretical Atomic and Molecular Physics\\
Harvard-Smithsonian Center for Astrophysics\\
Harvard, Cambridge, MA 02138\\
and\\
Joint Institute for Nuclear Research\\
141980, Dubna, Russia}
\maketitle

\begin{abstract}
The recently reported precise experimental determination of the dipole
polarizability of the $\mbox{H}_2^+$ molecular ion ground state
[P.L.~Jacobson, R.A.~Komara, W.G.~Sturrus, and S.R.~Lundeen, Phys.\ Rev.~A
{\bf62}, 012509 (2000)] reveals a discrepancy between theory and
experiment of about $0.0007a_0^3$, which has been attributed to
relativistic and QED effects. In present work we analyze an influence of
the relativistic effects on the scalar dipole polarizability of an
isolated $\mbox{H}_2^+$ molecular ion. Our conclusion is that it accounts
for only 1/5 of the measured discrepancy.
\end{abstract}

\section{Introduction}

Recent measurements \cite{exp98,Lundeen} of the scalar electric dipole
polarizability of $\mbox{H}_2^+$ molecular ion through the study of
$\mbox{H}_2$ molecule states with one Rydberg electron stimulated the
introduction of methods \cite{Shertzer}--\cite{Babb} which are able to
accurately describe wave functions of molecular ions with two heavy nuclei
beyond the adiabatic approximation. The accuracy for the dipole
polarizability constant ($\sim10^{-7}a_0^3$) reached in the last work
\cite{Babb} in its turn become a challenge to experiment. The new
experimental work \cite{Lundeen} substantially increases the accuracy of
measurements and reveals a discrepancy of about $0.0007a_0^3$ between
theory and experiment, which can not be accounted for within purely
nonrelativistic approximation. In present work we consider relativistic
corrections of order $\alpha^2$ to the dipole polarizability of the ground
state of an isolated $\mbox{H}_2^+$ molecular ion.

\section{Theory}

The nonrelativistic Hamiltonian of the hydrogen molecular ion $\mbox{H}_2^+$
is
\begin{equation} H_0=-\frac{1}{2M}\bn^2_1-\frac{1}{2M}\bn^2_2-\frac{1}{2m}\bn^2
+\frac{1}{R_{12}}-\frac{1}{r_1}-\frac{1}{r_2}, \end{equation}
We adopt atomic units ($e=\hbar=m=1$) throughout this paper. The interaction
with an external electric field (details of the nonrelativistic treatment of
the problem can be found in previous papers \cite{Shertzer}--\cite{Babb}) is
expressed by
\begin{equation} V_p = \mathcal{E}{\bf n}\cdot{\bf d}, \end{equation}
where
\[ {\bf d}=\mu{\bf r}_c=\left(\frac{2M}{2M+m}+2\frac{m}{2M+m}\right)
\left[{\bf r}-\frac{{\bf r}_1+{\bf r}_2}{2}\right] \]
is the electric dipole moment of the three particles with respect to the
center of mass of the system. Without loss of generality we assume that
${\mathbf{n}}\cdot{\mathbf{d}}=\mu z_c$.

The Breit $\alpha^2$ correction to the nonrelativistic Hamiltonian is
described by an operator
\begin{equation} V_B = \alpha^2\left\{-\frac{p^4}{8m^3}+\frac{4\pi}{8m^2}
\left[\delta({\bf r}_1)+\delta({\bf r}_2)\right]+
\frac{1}{2m^2}\left[\frac{[{\bf r}_1\times {\bf p}]}{r^3_1}+
\frac{[{\bf r}_2\times {\bf p}]}{r^3_2}\right]\frac{\bs}{2}\>\right\}.
\end{equation}
Then the total Hamiltonian reads,
\begin{equation} H=H_0+V_B+V_p\>. \end{equation}

Let us define the ground state nonrelativistic wave function as follows
\begin{equation} (H_0-E_0)\Psi_0=0. \end{equation}
In the nonrelativistic case the change of energy due to polarizability
of molecular ion is expressed by
\begin{equation} E_p^{(2)}=\langle\Psi_0|V_p(E_0-H_0)^{-1}V_p|\Psi_0\rangle=
{\mathcal{E}}^2\mu^2\langle\Psi_0|z_c(E_0-H_0)^{-1}z_c
|\Psi_0\rangle=-\frac{1}{2}\alpha_s^0{\mathcal E}^2, \end{equation}
and
\begin{equation}
\alpha_s^0=-2\mu^2\langle\Psi_0|z_c(E_0-H_0)^{-1}z_c|\Psi_0\rangle.
\end{equation}

Let us introduce $H_1=H_0+V_B$, then the scalar dipole polarizability
$\alpha_s$ with account of relativistic corrections can be rewritten in a
form (we assume that $V_B\approx\alpha^2 H_0$ and $\Psi_0^B=\Psi_0+\Psi^B$)
\begin{equation} \begin{array}{r@{}l}
\alpha_s^1&=-2\mu^2\langle\Psi_0^{B}|z_c(E_1-H_1)^{-1}z_c|\Psi_0^B\rangle\\[3mm]
&=-2\mu^2\langle\Psi_0^B|z_c\left[(E_0-H_0)^{-1}+
(E_0-H_0)^{-1}\left(V_B-\langle V_B\rangle\right)
(E_0-H_0)^{-1}+\dots\right]z_c|\Psi_0^B\rangle\\[3mm]
&=-2\mu^2\langle\Psi_0|z_c(E_0-H_0)^{-1}z_c|\Psi_0\rangle
-2\mu^2\langle\Psi_0|z_c(E_0-H_0)^{-1}V_B(E_0-H_0)^{-1}z_c|\Psi_0\rangle\\[2mm]
&\hspace{4mm}-2\mu^2
\Bigl(\langle\Psi^B|z_c(E_0-H_0)^{-1}z_c|\Psi_0\rangle
+\langle\Psi_0|z_c(E_0-H_0)^{-1}z_c|\Psi^B\rangle\Bigr),
\end{array} \end{equation}
and $\Psi^B=(E_0-H_0)^{-1}V_B|\Psi_0\rangle$.
Thus relativistic correction to the scalar dipole polarizability $\alpha_s$
is reduced to evaluation of the following matrix elements
\begin{equation}\label{relcor} \begin{array}{r@{}l}
\Delta\alpha_s=
&-2\mu^2\langle\Psi_0|z_c(E_0-H_0)^{-1}\left(V_B-\langle V_B\rangle\right)
 (E_0-H_0)^{-1}z_c|\Psi_0\rangle\\[3mm]
&-2\mu^2\langle\Psi_0|V_B(E_0-H_0)^{-1}z_c(E_0-H_0)^{-1}z_c|\Psi_0\rangle\\[3mm]
&-2\mu^2\langle\Psi_0|z_c(E_0-H_0)^{-1}z_c(E_0-H_0)^{-1}V_B|\Psi_0\rangle.
\end{array} \end{equation}
At this point we can note that the spin--orbit term does not contribute to
$\alpha_s$ since the magnetic dipole operator has selection rules $m'=m\pm1$.

\section{Variational nonrelativistic wave function}

Variational wave function describing the ground state of the hydrogen
molecular ion is taken in a form
\begin{equation} \begin{array}{r@{}l}
\displaystyle \Psi_0=&
\displaystyle \sum_{i=1}^{\infty}\Big[C_i\cos(\nu_i R_{12})+
D_i\sin(\nu_i R_{12})\Big]\\[2mm]
&\displaystyle \hspace{30mm} \times e^{-\alpha_i r_1-\beta_i r_2
-\gamma_i R_{12}}+({\mathbf r}_1 \leftrightarrow {\mathbf r}_2).
\end{array} \end{equation}
Here $\alpha_i$, $\beta_i$, $\gamma_i$, and $\nu_i$ are parameters generated
in a quasirandom manner,
\[
\alpha_i = \left\lfloor\frac{1}{2}i(i+1)\sqrt{p_\alpha}
\right\rfloor[(A_2-A_1)+A_1],
\]
where $\lfloor x\rfloor$ designates the fractional part of $x$, $p_\alpha$ and
$q_\alpha$ are some prime numbers, the end points of an interval $[A_1,A_2]$
are real variational parameters. Parameters $\beta_i$, $\gamma_i$, and
$\nu_i$ are obtained in a similar way. Details of the method and discussion
of various aspects of its application can be found in our previous papers
\cite{var99,var00}.

The perturbed function $\Psi_1=\mu(E_0-H_0)^{-1}{\bf r}_c\Psi_0$ is
expanded in the similar way
\begin{equation} \begin{array}{r@{}l}
\displaystyle \Psi_1=&
\displaystyle \sum_{i=1}^{\infty}{\bf r}_1\Big[\hat{C}_i\cos(\nu_i R_{12})+
\hat{D}_i\sin(\nu_i R_{12})\Big]\\[2mm]
&\displaystyle \hspace{30mm} \times e^{-\alpha_i r_1-\beta_i r_2
-\gamma_i R_{12}}+({\mathbf r}_1 \leftrightarrow {\mathbf r}_2).
\end{array} \end{equation}
Technically evaluation of the matrix elements in Eq.~(\ref{relcor}) can
proceed in the following way. For the matrix element in the first line we
need to solve one linear equation, $(E_0-H_0)\Psi_1=z_c\Psi_0$, and then
average operator $(V_B-\langle V_B\rangle$ over $\Psi_1$. To get rid of
singularities in the solutions of implicit equations of lines 2 and 3 of
Eq.~(\ref{relcor}) one can solve a sequence of equations from the right to
the left in the second line and in the reverse order for the third line. In
the latter two cases solution as well as the rhs of the last equation should
be projected onto subspace orthogonal to $\Psi_0$.

\section{Results and Conclusions}

In Table I results of numerical calculations are presented. For the
zeroth-order approximation a wave function with a basis set of $N=800~$ has
been used that yields the nonrelativistic energy
\begin{equation}
 E_0=-0.59713906312340(1),
\end{equation}
which is in a good agreement with our previous accurate result
\cite{var00}. Here $m_p=1836.152701m_e$ is adopted. As seen from the Table
convergence for the relativistic correction is slower due to singular
operators encountered in the matrix elements. Nevertheless we can
conclude from this Table that the resulting value is
\begin{equation}
\Delta\alpha_s = -0.00015214(1).
\end{equation}

Combining this result with the nonrelativistic value from the Table I one
obtains that the static electric dipole polarizability of $\mbox{H}_2^+$
molecular ion ground state with relativistic $\alpha^2$ corrections is to
be
\begin{equation}
\alpha_s = 3.1685737(1).
\end{equation}
We see that thus obtained value for the dipole polarizability does not
fully account for present disagreement between theory and experiment
(comparison of our results with results of previous calculations and
experiment are presented in Table II). Our consideration does not include
leading order QED corrections but usually they are one order of magnitude
smaller than relativistic corrections and have a different sign. Thus they
could not cover the rest 4/5 of the discrepancy.

On the other hand the experimental value for the dipole polarizability has
been deduced from the effective model Hamiltonian \cite{exp91} which is a
fully nonrelativistic Hamiltonian and it does not include the retardation
Casimir--Polder effect \cite{Casimir} for the Rydberg electron. On the
importance of this phenomena has been pointed out in a paper of Babb and
Spruch \cite{Babb-Spruch}. So, our conclusion is that the Casimir--Polder
potential has to be included into the effective Hamiltonian to meet the
requirements of the present level of experimental accuracy. That will enable
to deduce scalar electric dipole polarizability in a more reliable way.

This work was supported by the National Science Foundation through a grant
for the Institute for Theoretical Atomic and Molecular Physics at Harvard
University and Smithsonian Astrophysical Observatory, which is gratefully
acknowledged.

\clearpage

\begin{table}[h]
\begin{center}
\begin{tabular}{lll}
      & $~~~~\alpha_s^{0}$ & $\Delta\alpha_s$ \\\hline
 $N= 400$ & 3.1685962 & $-$1.52065[$-$4] \\
 $N= 600$ & 3.1687252 & $-$1.52140[$-$4] \\
 $N= 800$ & 3.1687258 & $-$1.52137[$-$4] \\
\end{tabular}
\end{center}
\caption{Dipole polarizability of $\mbox{H}_2^+(0,0)$.
Convergence of the numerical results with the size of basis set.}
\end{table}

\begin{table}[h]
\begin{center}
\begin{tabular}{ll}
      & $~~~~\alpha_s$ \\ \hline
 Experiment \cite{Lundeen}           & 3.16796(15) \\
 \\
 Shertzer and Greene \cite{Shertzer} & 3.1682(4)  \\
 Bhatia and Drachman \cite{Bhatia}   & 3.1680     \\
 Moss \cite{Moss}                    & 3.16850    \\
 Taylor, Dalgarno, Babb \cite{Babb}  & 3.1687256(1) \\
present work \\
~~~~nonrelativistic                  & 3.1687258  \\
~~~~with $\alpha^2$ corrections      & 3.1685737 \\
\end{tabular}
\end{center}
\caption{Dipole polarizability of $\mbox{H}_2^+(0,0)$. Comparison
with other calculations and experiment.}
\end{table}

\end{document}